%% file: accepted_manuscript.tex
\title{The complementarity of a diverse range of deep learning features extracted from video content for video recommendation}

\documentclass[review]{elsarticle}
\makeatletter
\def\ps@pprintTitle{%
 \let\@oddhead\@empty
 \let\@evenhead\@empty
 \def\@oddfoot{}%
 \let\@evenfoot\@oddfoot}
\makeatother

\RequirePackage{amsmath}
\RequirePackage{amsfonts}
\RequirePackage{amstext}
\RequirePackage{amssymb}
\RequirePackage{mathtools}
\RequirePackage{bm}
\RequirePackage[mathcal]{eucal}
\RequirePackage{upgreek}

\usepackage{fontawesome5}

\usepackage{lineno,hyperref}
\usepackage[nolist, nohyperlinks]{acronym}
\usepackage{hyperref}
\usepackage{enumitem}
\usepackage{amssymb}
\usepackage{fancyhdr}
\usepackage{longtable}
\usepackage{adjustbox}
\usepackage{booktabs,caption}
\usepackage{multirow}
\usepackage{ltablex}
\usepackage{pdflscape}
\usepackage{xcolor}
\usepackage{graphicx}
\usepackage{stackengine}
\usepackage{longtable}
\usepackage{float}
\DeclarePairedDelimiter\floor{\lfloor}{\rfloor}
\def\boxit#1{%
  \smash{\fboxsep=0pt\llap{\rlap{\fbox{\strut\makebox[#1]{}}}~}}\ignorespaces
}
\modulolinenumbers[5]

\journal{Expert Systems with Applications}

\biboptions{authoryear}

\begin{document}
\input{acro_list}

\begin{frontmatter}









\title{The complementarity of a diverse range of deep learning features extracted from video content for video recommendation}

\author[university]{Adolfo Almeida \corref{cor1}}
\ead{u13010396@tuks.co.za}

\author[university]{Johan Pieter de Villiers}
\ead{pieter.devilliers@up.ac.za}

\author[university]{Allan De Freitas}
\ead{allan.defreitas@up.co.za}

\author[multichoice]{Mergandran Velayudan}
\ead{mergandran.velayudan@multichoice.co.za}

\cortext[cor1]{Corresponding author.}
\address[university]{Department of Electrical, Electronic and Computer Engineering, University of Pretoria, Private Bag X20, Hatfield 0028, South Africa}
\address[multichoice]{AI Centre of Excellence, MultiChoice Group, PO Box 1502, Randburg 2125, South Africa}


\begin{abstract}
Following the popularisation of media streaming, a number of video streaming services are continuously buying new video content to mine the potential profit from them. As such, the newly added content has to be handled well to be recommended to suitable users. In this paper, we address the new item cold-start problem by exploring the potential of various deep learning features to provide video recommendations. The deep learning features investigated include features that capture the visual-appearance, audio and motion information from video content. We also explore different fusion methods to evaluate how well these feature modalities can be combined to fully exploit the complementary information captured by them. Experiments on a real-world video dataset for movie recommendations show that deep learning features outperform hand-crafted features. In particular, recommendations generated with deep learning audio features and action-centric deep learning features are superior to MFCC and state-of-the-art iDT features. In addition, the combination of various deep learning features with hand-crafted features and textual metadata yields significant improvement in recommendations compared to combining only the former.             
\end{abstract}


\end{frontmatter}


\section{Introduction}
\insert\footins{
  \normalfont\footnotesize
  \interlinepenalty\interfootnotelinepenalty
  \splittopskip\footnotesep \splitmaxdepth \dp\strutbox
  \floatingpenalty10000 \hsize\columnwidth
  \noindent This is the accepted manuscript version of this article, which has been published in Expert Systems With Applications international journal \href{https://doi.org/10.1016/j.eswa.2021.116335}{https://doi.org/10.1016/j.eswa.2021.116335} © 2021. This work is licensed under a \href{https://creativecommons.org/licenses/by-nc-nd/4.0/}{CC-BY-NC-ND 4.0} International license
  \faCreativeCommons\ \faCreativeCommonsBy\ \faCreativeCommonsNc\ \faCreativeCommonsNd\par}
As video streaming platforms become more prevalent in our society, large amounts of video data are increasingly being uploaded to video sharing websites~\citep{Frahm_2017_CVPR_Workshops}. The video sharing websites depend heavily on video recommendation systems to assist users to discover videos they may enjoy. A video recommendation system is a user-level video filtering service which helps users explore the world of videos \citep{adomavicius2005toward}. It offers a more personalised experience to users by recommending the most relevant and appropriate videos for them. In order to do this, algorithms are used to analyse the information about the videos, users and past interactions between them \citep{gomez2016netflix,lu2015recommender}.

Existing recommendation systems mainly use one of three approaches, namely the \ac{CF} recommendation method, the \ac{CB} recommendation method, and the hybrid recommendation method, which is a combination of the two former recommendation approaches \citep{adomavicius2005toward}. The \ac{CF} recommendation method uses the user's explicit or implicit feedback, such as previous ratings and watch history in order to predict the preference of a user. This is achieved by recommending a video to a user if like-minded users have given it a positive rating or have watched it \citep{adomavicius2005toward}. The \ac{CB} recommendation method uses the target user's profile and video content to predict the target user preferences. A video is recommended to a user if its content is similar to what the user liked or watched before \citep{lops2011content}. On the other hand, the hybrid recommendation methods combine both the user's feedback and the consumed video content in order to improve recommendations. 

Most video streaming services that use a video recommendation system to compute the video relevance based on user feedback \citep{liu2018content} use \ac{CF} recommendation methods because of their state-of-the-art accuracy \citep{deldjoo2019movie,yuan2016solving}. This feedback is used to model the user-video preference and compute video-to-video relevance scores in order to provide personalised recommendations. However, this approach suffers from the new item cold-start problem \citep{wei2017collaborative, deldjoo2019movie}. This problem is a core problem in the recommendation field \citep{wang2019overview,elahi2019user,volkovs2017dropoutnet}. It is a serious problem faced by video streaming services that purchase new movies and TV series from content providers \citep{liu2018content,wang2019overview}. Moreover, with the tremendous increase in the number of new videos being continuously uploaded, some video streaming services have to deal with unrated, unaudited and completely new content of which they do not know anything about \citep{kumar2018icebreaker}. As such the new item cold-start problem has to be handled well in order for the uploaded content to be discovered by most of their users.

In addition, because of the massive amount of videos being produced, it is unfeasible to rely on manual processing of multimedia data to solve a wide variety of multimedia problems \citep{shen2020advance}. As a result, recent studies on video content analysis and specially video retrieval tasks use various types of deep learning features extracted using pre-trained models due to their outstanding performance in different domains compared to hand-crafted features \citep{shen2020advance, tran2015learning, liu2019use, miech2019howto100m}. Furthermore, deep learning features also require fewer pre-processing steps compared to traditional methods \citep{shen2020advance}. Hence, it is a practical solution for a vast number of tasks, particularly when dealing with large-scale video datasets. However, in the field of video recommendation, the utilisation of several deep learning features that capture different aspects of the video content is still a rare, explored area compared to hand-crafted features~\citep{deldjoo2020recommender}.

Recent work on personalised video recommendation for video streaming services \citep{deldjoo2019movie,du2018personalized} have shown that video recommendation based on deep learning object features and hand-crafted features combined with collaborative filtering information have a higher recommendation quality compared to recommendations based on only deep learning object features or metadata such as genre or cast. However, in order to solve the new item cold-start problem, they either combine only two features \citep{deldjoo2019movie}, i.e., deep learning object features and hand-crafted audio features; or combine deep learning object features along with hand-crafted motion and audio features \citep{du2018personalized}; or combine only deep learning object features and genre features \citep{ma2018lga}. In addition, they limit themselves by not exploring deep learning action features which captures the motion information in the videos and their complementariness among deep learning visual-appearance and audio features. This information is important since it is part of the rich and varied additional multimodal information present in videos. Videos are characterised by actions and scenes that help its narrative and pass on their message to the audience~\citep{carreira2017quo,huang2018makes,stroud2018d3d,adeli2019component,wehrmann2017movie}, which may influence the users' preferences to a considerable extent. For example, temporal sequencing of cars in a video where the cars in the scene might appear stationary, yet the background is continually moving could be an indicative of a car chase; an irregular and complex kind of motion could be an indicative of hand-held shot videos which some people do not like~\citep{alvarez2019influence}.

It is clearly evident that there is a need to solve the new item cold-start problem by implementing a video recommendation system that uses the users' preference history and considers the complementary information among different deep learning features extracted from the video content. These features should capture the visual-appearance, audio and motion information from the video content in order to best exploit their availability and provide more accurate personalised video recommendation to users in the new item cold-start scenario. 

In this paper, we address the new item cold start problem by enhancing the recommendation task using visual-appearance, audio and action deep learning features to recommend newly added videos to users effectively. In particular, the performance of these features is evaluated in terms of accuracy and beyond-accuracy metrics in the item warm-start and cold-start scenarios. We compare the deep learning features against genre features and hand-crafted features. In addition, we investigate fusion methods to exploit the complementary information captured by the deep learning features and enrich the quality of recommendation. Finally, we also perform an ablation study to empirically assess the importance of using a diverse range of video content features on the overall recommendation quality while taking full advantage of the available data. Therefore, this experiment is conducted by combining the video content features evaluated in this research study. The major contributions of this work are as follows:

\begin{enumerate}
    \item  We evaluate the performance of various state-of-the-art deep learning features that capture object, scene, audio and action information from video content for item warm-start and cold-start video recommendations. These features are more versatile compared to hand-crafted features and metadata since they are not computationally expensive, require fewer pre-processing steps and only raw video content, and metadata may not be precise or available.   
    
    \item We compare the deep learning features to genre metadata, \ac{MoSIFT} features, \ac{MFCC} features and state-of-the-art \ac{iDT} features in terms of accuracy and beyond-accuracy metrics. 
    
    \item We evaluate and compare the performance of different early fusion techniques to determine how well each fusion method combines the different deep learning features that capture visual-appearance, audio and motion information contained in the videos in order to enrich the quality of recommendations.
    
    \item We take full advantage of the available data by combining the video content features explored in this work. The importance of each video content feature on the overall recommendation quality is assessed. To the best of our knowledge, this is the first study to combine this diverse range of video content features, particularly deep learning features, in order to alleviate the new item cold start problem; and where each feature is removed to see how that affects the overall performance of the recommendation system. 
    
    \item We propose an improved \ac{CER} model that uses a matrix scaling technique to enhance the performance of the \ac{CER} model in both item warm start scenario and item cold start scenario.  
\end{enumerate}

The remaining sections of this paper are organised as follows: Section \ref{sec:approach} describes the video content features, the video recommendation model and provides details of the feature fusion methods used in this study. In Section \ref{sec:experiments}, the experimental settings and evaluation results of the investigation are discussed. Lastly, Section \ref{sec:conclusion} presents the conclusions of the proposed work and future work.

\section{Approach}\label{sec:approach}
\begin{figure}[H]
\centering

\includegraphics[width=1.\linewidth]{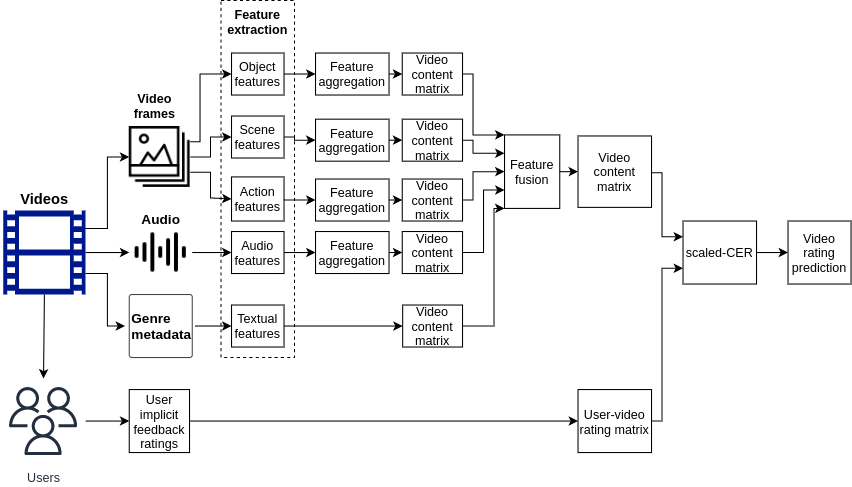}
\captionof{figure}{General overview of the proposed workflow used to generate video recommendations. Given a set of videos watched by a set of users, a diverse range of deep learning features are first extracted from the visual and audio modalities of each video, and textual features are obtained from genre metadata. Next, the deep learning features of each video are temporally aggregated to obtain a video content vector of fixed length, which represents the information about their content. These vectors are then combined to form a video content matrix where each row represents a single video. Finally, the video content matrices are fused and used with the user-video rating matrix as input to the scaled-CER model to train it and then generate recommendations.}
\label{recsys}

\end{figure}

In this work, \textit{it is assumed that all the movie trailers are available when training the recommender model and the visual and audio features from movie trailers are representative of the features extracted from complete films} \citep{deldjoo2019movie, deldjoo2016content}. For this reason, deep learning features are extracted from movie trailers instead of the complete movie videos in this research work. This allows the exploration of the multimodal information from video content to be computationally efficient. Figure \ref{recsys} shows the general overview of the proposed workflow used to generate video recommendations. The time complexity of the feature extraction phase of this workflow is proportional to the total number of videos and their duration, and the time complexity of recommender model is proportional to the total number of features which represent the videos. 

\subsection{Deep learning features}
The deep learning features used in this work are \ac{CNN} embeddings generated by intermediate layers of \ac{CNN} models. These embeddings are chosen because they are more generalised and robust to noise opposed to features extracted from the final output layer \citep{holzenberger2019learning, kalliatakis2019exploring}. The visual-appearance information contained in the videos is represented by object-centric and scene-centric \ac{CNN} embeddings. The motion information is represented by action-centric \ac{CNN} embeddings. Lastly, the audio information is represented by audio \ac{CNN} embeddings. The feature extraction process is described below for each visual-appearance, audio and action deep learning features.

\begin{enumerate}[label = \Alph*.]

\item \textbf{Object features\\}
The object information from videos is captured using a \textit{Obj(IN)} model pre-trained on the ImageNet dataset \citep{deng2009imagenet} for the task of object classification. This model is a ResNet-152 network \citep{he2016deep} that receives as input images of size $224\times224$ pixels and 3 channels. Therefore, the videos are decoded at 1 frame per second (fps) and each video frame is resized to $224\times224$ pixels. The object-centric embeddings are extracted from the last convolutional layer with 2048 neurons followed by a global spatial average pooling layer. Thus, each video frame is represented by a 2048-dimensional descriptor which contains video content object features present in the frame. As a result, the output dimension of the \textit{Obj(IN)} feature extractor for each video decoded at 1 fps is $F \times 2048$ where $F$ is the total number of frames, i.e. a 120 second video is represented by a descriptor of size $120 \times 2048$.

\item \textbf{Scene features\\}
The scene where an action is taking place may provide relevant information that supports actions with object interactions. In this work, the scene information from video frames is captured using a DenseNet-161 model \citep{huang2017densely} pre-trained on the Places365 dataset \citep{zhou2017places}. This model is a 2D-\ac{CNN} network with 161 layers. It consists of an input layer that receives $224\times224$ input images. Thus, each frame is first resized to this scale before it is passed to the model. The scene-centric embeddings are extracted from the $7\times7$ global average pooling layer of the DenseNet-161 model \citep{huang2017densely} followed by a global spatial average pooling layer resulting in a 2208-dimensional descriptor for each video frame. This descriptor contains video content scene features that represent related contextual information about a scene in a frame. As a result, similar to the object feature extractor, the scene-centric embeddings are extracted from  videos  decoded  at  1 fps. The output dimension of the scene feature extractor is $F \times 2208$ where $F$ is the total number of frames, i.e. a 120 second video is represented by a descriptor of size $120 \times 2208$.  

\item \textbf{Action features\\}
The action features are extracted from videos with pre-trained 3D-\ac{CNN} models. These features capture the motion information in a video \citep{carreira2017quo}. In particular, each video is decoded at 24 fps and the visual stream is used as input to \textit{Action(IG)} and \textit{Action(HMDB)} models.  

The \textit{Action(IG)} is a R(2+1)D-34 32-frames model \citep{tran2018closer} pre-trained on the IG-65m dataset \citep{ghadiyaram2019large} that includes 359 human action classes that are identical to the action labels of the Kinetics dataset \citep{carreira2017quo}. This model consists of 34 layers where 33 layers are convolutional layers and the final layer is a fully-connected layer with softmax (classification layer). The input layer receives clips consisting of 32 consecutive RGB video frames with size $112\times112$ pixels. Thus, if the clips obtained from the video are composed of video frames with different resolutions, these frames need to be resized. The last convolutional layer has 512 neurons. The embeddings generated by this layer are passed to a global spatio-temporal average pooling layer and fed into a fully-connected layer with classification layer that predicts the 359 action classes. The action-centric embeddings generated by the global spatio-temporal average pooling layer are extracted and used as the video-clip content action features. They form a 512-dimensional descriptor for each clip of 32 consecutive $112\times112$ pixel frames. Having said that, the output dimension of the \textit{Action(IG)} feature extractor is $Tv \times 512$ where $Tv = \frac{F}{32}$, i.e. a 120 second video decoded at 24 fps has 2880 frames. As a result, this video is represented by a descriptor of size $90 \times 512$ since $Tv = \frac{2880}{32} = 90$.

The \textit{Action(HMDB)} model is a ResNeXT-101 64-frames network \citep{hara2018can} pre-trained on the Kinetics dataset \citep{carreira2017quo} and fine-tuned on the HMDB-51 dataset \citep{kuehne2011hmdb}. This model is chosen because it has been trained to recognise sequences of actions from untrimmed digitised movies. It consists of 101 layers where the last convolutional layer is followed by a global average pooling layer and a fully-connected layer with classification layer. The classification layer of the \textit{Action(HMDB)} model creates a distribution for the 51 labelled classes. The size of the input layer is $3 \textit{ channels}\times64 \textit{ frames}\times112 \textit{ pixels}\times112 \textit{ pixels}$. Therefore, a 64-frame clip needs to be resized if it has a different resolution. In this work, video frames are decoded at 24fps and processed in clips of 64 consecutive frames. Hence, every single clip spans approximately 2.67 seconds of the video. Each frame is first resized to $112\times112$ pixels, before passing to the model. Similar to \citep{almeida2020visual}, we extract the action-centric embeddings generated by the global average pooling layer that comes before the classification layer and yields a 2048-dimensional descriptor for each video-clip. These descriptors are taken as the action features. Therefore, in a similar fashion as the \textit{Action(IG)} feature extractor, the output dimension of the \textit{Action(HMDB)} feature extractor is  $Tv \times 2048$ where $Tv = \frac{F}{64}$, i.e. a 120 second video decoded at 24 fps is represented by a descriptor of size $45 \times 2048$.

\item \textbf{Audio features\\}
Audio features are extracted from audio frames with a \textit{VGGish} model \citep{hershey2017cnn}. This model is a 2D-\ac{CNN} network pre-trained on the YouTube-8m dataset for audio classification \citep{hershey2017cnn}. The model is a modified \textit{VGG} architecture. 

In order to extract sound features using this model the audio stream of each video needed to be pre-processed. The raw audio waveform is first downsampled to a 16 kHz mono signal with 16 bit resolution and re-scaled to the range [-1.0, 1.0]. Next, the audio signal is divided into a sequence of successive non-overlapping 0.96 sec audio segments of the original video, and subsequently converted from time domain to frequency domain. The conversion is performed with \ac{STFT}. This operation is computed using a periodic Hann window that receives as input frames with size of 25 ms and stride of 10 ms. The resulting spectrogram is mapped to 64 log Mel-spectrogram bins which in turn gives patches of $96$ \textit{audio-frames}$\times 64$ bins. These log Mel-spectogram patches form the input to the \textit{VGGish} model that maps them to a 128-dimensional descriptor for each audio segment. As a result, each 128-dimensional descriptor composed of \textit{VGGish} features represents 96 \textit{audio-frames}. For this reason, the dimension of the audio descriptor extracted for each video's audio track is $Ta \times 128$ \textit{VGGish} features, where $Ta = \floor*{\frac{\mathit{audioFrames}}{96}}$.

\end{enumerate}

As the number of frames varies across videos, the deep learning features are aggregated into video-level feature vectors using six statistical feature aggregation methods, namely maximum, mean, median, variance, median absolute deviation and interquartile range \citep{almeida2020visual}. These methods are chosen because they are simple and widely used on a number of video understanding tasks which utilise deep learning features \citep{liu2019use, miech2019howto100m, almeida2020visual} and obtained better results compared to state-of-the-art \ac{FV} and \ac{VLAD} aggregation methods \citep{abu2016youtube,deldjoo2019movie}. 

Additionally, to further enhance the discrimination of the video-level feature vectors, we use the \ac{SSR} normalisation followed by \ac{PCA} on the raw features. \ac{SSR} is executed in order to weaken the dominant dimensions of each video-level feature vector so they do not overshadow the other dimensions during the similarity computations \citep{du2018personalized}. This normalisation function is defined as 

\begin{equation}
    SSR(\mathbf{x}) = sign(\mathbf{x}) \times \sqrt{|\mathbf{x}|},
\end{equation}

\noindent where $\mathbf{x}$ is the video-level feature vectors and $sign()$ is the function that captures the sign of each feature. Moreover, \ac{PCA} is applied to obtain features that are more discriminative and less redundant. Hence, the number of principal components is equal to the original list of features in order not to lose any information while covering maximum variance among them. Furthermore, each video-level feature vector is scaled into a unit vector by applying $L_2$-normalisation ($L_2$-norm) given by Equation (\ref{eq:l2_norm}) below  

\begin{equation}
 L_2\mbox{-norm}(\mathbf{x}) = \frac{\mathbf{x}}{||\mathbf{x}||_2},
 \label{eq:l2_norm}
\end{equation}

\noindent where $||\mathbf{x}||_2$ is the Euclidean norm of the video-level feature vector defined as $||\mathbf{x}||_2 = \sqrt{\sum_{i=1}^{N}(x_i)^2}$. This is performed to ensure that each feature contributes approximately equally to the final similarity measure \citep{ranjan2017l2}. 

\subsection{Textual features}
Aside from the video features extracted from the video content, textual metadata features provide a good representation of the videos. Textual feature modality is the most used video representation in traditional \ac{CB} or hybrid approaches for video recommendation. Although the main of the objective of this research work is to investigate the effect of various visual and audio stimuli on user preferences, it is worth exploiting textual features as complementary information of video description. 

A set of genres of each movie are used as the only type of textual feature in this work. The motivation behind this choice is that genre metadata is highly available in the domain and represent relevant elements in movies \citep{deldjoo2019movie}. In addition, taking into account that they are high-level semantics attributes of movies, when fused with non-textual content features they will probably remove ambiguity which in turn should lead to an improvement in performance. 

Given the genres provided in the meta information of the videos, the genre feature vector is encoded to a $N$-dimensional binary vector where $N$ is the total number of unique genres. A bit $1$ in the $i$th column of the vector indicates that the corresponding genre describes the video and a bit $0$ indicates that the corresponding genre does not apply to the video. 

The genre feature vector used in this work represent 19 genre labels from the metadata of the movies, namely \textit{adventure, animation, children, comedy, fantasy, romance, drama, action, crime, thriller, horror, sci-fi, mystery, IMAX, documentary, war, film-Noir, musical}, and \textit{western}. Thus, the dimensionality of the genre feature vector for each video is 19 where each feature represents one of the 19 annotated genres.

\subsection{Video recommendation model}
The objective of this work is to explore different features that capture the rich and diverse multimodal information present in videos, which may influence the users’ preferences to a considerable extent \citep{deldjoo2020recommender} thereby alleviate the new item cold start problem. For this reason, the state-of-the-art \ac{CER} model \citep{du2018personalized} is chosen since it is a hybrid recommender model that could lead to the best benefit in terms of recommendation performance in item warm-start and cold-start scenarios using the wide variety of features. 

The \ac{CER} model is a model based on the weighted matrix factorisation method for implicit feedback datasets where a large matrix is decomposed into smaller matrices to reduce the dimensions and learn latent vectors that describe users and items. The implicit feedback ratings are turned into confidence values as follows \citep{du2018personalized}:

\begin{equation}
    c_{ui} = \begin{cases}
    1, & \text{if $r_{ui} = 1$},\\
    0.01, & \text{if $r_{ui} = 0$}.
  \end{cases}
\end{equation}

\noindent where $c_{ui}$ is the confidence value for the user-video pair $(u,i)$ given its rating $r_{ui}$ obtained from the \ac{URM}. The confidence values are used to learn the users and items latent vectors by performing the alternating least squares approach \citep{du2018personalized}. In addition, the CER model leverages the collaborative information from warm items with single type of video content features to effectively recommend warm and cold items. These features should be aggregated into video-level feature vectors that describe the video content of each video. 

Latent vectors are composed of latent factors which represent categories that are present in the data in a much lower dimensional space. These vectors are used by the \ac{CER} model to predict ratings that are missing in the original \ac{URM} since every user has videos that they have not watched before. These videos are recommended according to CER's rating predictor \citep{du2018personalized}. 

In this work, the CER model is trained using the optimal hyper-parameter set reported in the original paper \citep{du2018personalized}. However, the original paper does not mention the number of epochs and the stopping criteria used in the training step of the CER model. Therefore, in this work, the number of epochs is selected using the early stopping technique \citep{dacrema2019troubling}. This method decreases the risk of over-fitting and also decreases training time.

The main limitation of the \ac{CER} model is that it does not learn from multiple types of video content features at once. Therefore, there is a need to further investigate fusion methods to leverage the complementary information from the diverse range of features explored in this work in order to further enrich the recommendations. In addition, according to existing works \citep{lee2017large, ma2018lga}, the combination of video-level feature vectors should lead to an even higher recommendation quality in the new item cold-start scenario in contrast to the use of a single feature modality.

\subsubsection{Improving CER model using matrix scaling}\label{sec:scaledCER}

Recently, successful recommender models named \textit{EigenRec} \citep{nikolakopoulos2019eigenrec} and \sloppy \textit{hybridSVD} \citep{frolov2019hybridsvd} have shown significant recommendation quality improvement using a simple scaling trick. These models are matrix factorisation recommendation algorithms that apply \ac{SVD}. The scaling trick used by these models is a matrix scaling technique which regulates how the popularity of items affects the predicted ratings. It is defined as \citep{nikolakopoulos2019eigenrec}  

\begin{equation}
    \widetilde{\mathbf{R}} \triangleq \mathbf{R}\mathbf{D}^{d-1},
    \label{eq:matrix_scaling}
\end{equation}

\noindent where $\mathbf{R}$ is the $URM$, $\mathbf{D} = diag\{||\mathbf{r}_1||, ||\mathbf{r}_2||, \dots, ||\mathbf{r}_m||\}$ is a diagonal matrix that contains Euclidean norm scaling for a given scaling factor $d$ of the columns $\mathbf{r}_i$ of $\mathbf{R}$ and lastly $\widetilde{\mathbf{R}}$ is the modified $URM$. 

From Equation (\ref{eq:matrix_scaling}) above, it can seen that when $d$ is 1, the standard model is obtained (\ac{URM} is not modified). However, when the scaling factor is varied, the sensitivity of the SVD based models to the popularity of the items is modified. Higher values of the parameter $d$ increase the sensitivity to popular items while smaller values increase the sensitivity to rare items. This adjustment leads to a new model with a latent space with different internal structure. It has been found that values slightly below 1 yield the best recommendation performance for \textit{EigenRec} and \textit{hybridSVD} models \citep{nikolakopoulos2019eigenrec,frolov2019hybridsvd}. 

Therefore, enlightened by these new findings, we propose an improved CER model that uses the matrix scaling technique to enhance the performance of the CER model \citep{du2018personalized}. The matrix scaling technique is used to produce a scaled-CER model\footnote{\href{https://github.com/Adolfo-Almeida/scaled_CER}{https://github.com/Adolfo-Almeida/scaled\_CER}} which has an increased sensitivity to unpopular and new items in contrast to the original non-scaled CER model. This choice is also supported given the fact that a value of 1 for the parameter $d$ leads to the original non-scaled CER model. As a result, this indicates that the original non-scaled CER model implicitly chose this value that could lead to a model biased towards popular items. Therefore, many items that have only few ratings, unpopular or new items, might not be leveraged accordingly during the training process. Consequently, this implicit default choice could inevitably hinders the CER model potential in both item warm-start and cold-start scenarios. In this work, using the matrix scaling technique, the confidence parameter $c_{ui}$ for the user-video pair $(u,i)$ of the scaled-CER model is defined as 

\begin{equation}
    c_{ui} = \begin{cases}
    1 \times ||\mathbf{r}_i||^{d-1}, & \text{if $r_{ui} = 1$},\\
    0.01, & \text{if $r_{ui} = 0$}.
  \end{cases}
\end{equation}

The optimal scaling factor hyper-parameter $d$ is searched by optimising the quality of the scaled-CER in terms of $MAP@5$ on the validation set. This measure is defined in section \ref{sec:accuracy_metrics}. The hyper-parameter optimisation is conducted on all cross-validation folds individually with Bayesian optimisation \citep{deldjoo2019movie}. Bayesian optimisation selects the next set of hyper-parameters based on the results of the hyper-parameter sets previously evaluated. Once the optimal scaling factor is found on each CV fold, a single optimal scaling factor is selected corresponding to the best average $MAP@5$ result across all folds. Recent studies \citep{deldjoo2019movie,dacrema2019we} have shown that Bayesian optimisation is an efficient method for hyper-parameter tuning. The benefits of this method are a reduction in search time and better parameter values compared to random search or grid search parameter optimisation methods.

\subsection{Feature fusion}\label{sec:fusionAlgorithm}
Multimodal fusion can be a very important component in video recommendation systems where improving the overall recommendation quality of the system is considered as one of its most essential aspects. The feature fusion methods commonly used are late fusion and early fusion. Late fusion combines prediction scores of each model in order to obtain a more accurate final set of results. As a result, the main disadvantage of this method is the loss of complementary information represented by different features. This information is important for the final estimation. In addition, late fusion is computationally more expensive given the fact that it requires separate systems and a learning stage for the combination \citep{deldjoo2019movie}. 

On the other hand, early fusion obtains a truly multimedia feature representation. It exploits the complementary information about various characteristics of a video at feature level. This in turn improves the discriminativity of the video representations. In contrast to the late fusion approach, the early fusion approach only needs a single model and one learning stage. The video information is represented by features from different modalities, namely visual, aural, and textual that are combined into a single feature vector, before being fed to a machine learning algorithm. One recent work in video retrieval tasks shows that early fusion of object, action, face, audio, scene, optical character recognition, and text features allows the system to obtain a better similarity measure and therefore be capable of more robust video retrieval \citep{liu2019use}. An increase on the overall performance of the system is observed when different features are cumulatively fused \citep{liu2019use}. 

Inspired by the above-mentioned findings, various early fusion methods are investigated to enrich the recommendations. This is executed to fully exploit the complementary information from the various feature representations extracted from the video content. Furthermore, this is also to determine whether early fusion would achieve a similar outcome observed in recent video retrieval tasks \citep{liu2019use,miech2019howto100m}. We hypothesise that a video recommendation system that uses videos represented in a shared unified space by diverse deep learning features (visual-appearance, audio and motion) should further improve the quality of recommendations in the new item cold-start scenario. %

The early fusion approaches, investigated to combine information from multiple modalities are the \ac{concat} method, the \ac{sum} method and lastly the \ac{max} method \citep{kalliatakis2019exploring}. The concat method is a technique that merges different feature vectors to obtain one large feature vector that represents the final video representation. As this feature vector contains many features, it increases the training time. Formally, for each video feature vector $\mathbf{f}$, if there are $L$ feature vectors of different modalities that are represented with $f_i \in \mathbb{R}^{d_i}$, the concatenation operation is defined as

\begin{equation}
    \mathbf{x} = [\mathbf{f}_1, \mathbf{f}_2, \dots, \mathbf{f}_L],
\end{equation}

\noindent where $\mathbf{x}$ is the final multimodal video-level representation by fusing the different features that capture visual-appearance, audio and motion information from videos as well as textual information from their metadata. The final size of this representation is the sum of the dimensions of all feature vectors denoted as $d = \sum_{i=1}^{L}d_i$.

The second early fusion method exploited in this work is the sum method. This method adds different feature vectors in order to obtain the final video representation. Given a set of $L$ feature vectors with the same size that represent each video modality separately, their summation is denoted as

\begin{equation}
    \mathbf{x} = \sum_{i=1}^{L} \mathbf{f}_i,
    \label{eq:sum_fusion}
\end{equation}

\noindent where $\mathbf{x}$ is the final multimodal representation with size $d$. As can be seen in Equation (\ref{eq:sum_fusion}), the sum fusion technique is only defined if all feature vectors have the same size. For cases where a feature vector $i$ of size $d_i$ is greater than $min(d_1, d_2, \dots, d_i, d_L)$, \ac{PCA} is applied for feature reduction. The number of features is reduced to the size of the smallest feature vector before performing the fusion operation. 

Finally, the last fusion technique investigated is the max fusion operation. This fusion method is similar to the sum fusion method in terms of the final multimodal representation size, however, it differs in the way the feature vectors are combined. The max fusion method selects the highest value of each feature from a set of $L$ feature vectors with the same size as 
\begin{equation}
    \mathbf{x} = max_{i=1}^{d}(f_1^{i}, f_2^{i}, \dots, f_L^{i}),
\end{equation}
where $\mathbf{x}$ is the final video representation and $d$ is the feature vector size. Similar to the sum fusion method, \ac{PCA} is applied as a dimension reduction step for all feature vectors greater than the smallest feature dimension in the set of $L$ feature vectors.

\section{Experimental results and discussion}\label{sec:experiments}
In this section, we present the dataset, evaluation metrics, experimental results, and the in-depth analyses of the reported results. The video recommendation system is investigated in the item warm-start and cold-start scenarios. The item warm-start scenario represents the case when some preference data for items in that scenario have been used to train the video recommendation model. The new item cold-start scenario represents the case when preference data for the items in that scenario are not known at all during the training.

\subsection{Dataset Description}\label{sec:dataset}

In this work, we use a processed MovieLens-10M dataset \citep{du2018personalized}. This dataset is a processed version of the publicly available MovieLens-10M dataset where 10380 movie trailers out of the 10682 movies in the MovieLens-10 dataset are downloaded from YouTube and manually checked if they are correct. The movies that have their trailers missing are removed. The data are binarized by converting ratings of 5 to 1, and all other ratings to 0. The processed dataset also provides five cross-validation folds where each fold is divided into a training set, a item warm-start test set and a new item cold-start test set. These sets contain the binarized ratings of 69878 users for each 10380 movie and occupy 60\%, 20\% and 20\%, respectively \citep{du2018personalized}. The item warm-start and cold-start test sets correspond to the item warm-start and cold-start scenarios where the item warm-start test set contains items that have some of their ratings in the training set and the item cold-start test set contains items that do not have any ratings in the training set. In addition, the dataset also provides trailers of movies and pre-computed 4000-dimensional \ac{MFCC} \citep{du2018personalized}, \ac{MoSIFT} \citep{chen2009mosift} and \ac{iDT} \citep{wang2013action} feature vectors for each movie trailer. These hand-crafted features are used as the video content feature baselines.

\subsection{Evaluation metrics}
The performance of the video recommendation system is evaluated using two types of metrics namely: accuracy and beyond-accuracy metrics \citep{deldjoo2019movie, shani2011evaluating}. These metrics are important when evaluating a video recommendation system since they complement each other. Accuracy metrics evaluate the relevance of the recommendations, however better user satisfaction beyond relevance is not necessarily achieved with higher accuracy \citep{silveira2019good}. Beyond-accuracy metrics evaluate the value that recommendations can generate to the user where the desire for variety is not ignored \citep{shani2011evaluating}.  

In this work, the videos in the catalogue are sorted in descending order, based on the ratings estimated by the model being evaluated. Next, the Top-$n$ videos are chosen to be the first $n$ videos in the recommendation list. The length $n$ of the recommendation list returned to each user is also known as the cut-off value. Accuracy and beyond-accuracy metrics are calculated for 3 different cut-off values from \{5, 15, 30\}.  All reported results for the recommendation models evaluated in the item warm-start and cold-start scenarios are obtained using the item warm-start test set and the item cold-start test set, respectively. The experiments were conducted using 5-fold cross-validation with the provided five-folds. Therefore, the presented results is the mean of five tests.

\subsubsection{Accuracy metrics}\label{sec:accuracy_metrics}
In order to evaluate if the user enjoyed the videos recommended by the video recommendation system using the various deep learning features, the following rank-aware metrics are used as listed below:\\

\begin{enumerate}
    
    \item \ac{MAP} is the average of the \textit{average precision at top N recommendations} $(AP@N)$ over the whole set of users in the test set. It calculates the overall precision of the recommender system by using precision at all possible recall levels \citep{deldjoo2019movie}. $AP@N$ is computed by obtaining the arithmetic mean of precision values of the relevant items at their corresponding positions. This metric is chosen because it measures the rate of relevant items in the recommendation list that users may like and therefore consumed, while considering relevant items not in the recommendation list. It is an important metric if it is assumed that many users will not scan the entire recommendation list, but instead they would only look at the top of the recommendation list. This metric is defined as \citep{deldjoo2019movie} 
    
    \begin{equation}
        AP_u@N = \frac{1}{min(N,K)} \sum^{N}_{i=1}P@i \cdot rel(i) ,
        \label{AP}
    \end{equation}
    
    \begin{equation}
        MAP = \frac{1}{|U|}\sum_{u \epsilon |U|}AP_{u},
    \end{equation}
    
    where $N$ is the length of the recommendation list, $K$ is the total number of relevant items, $P@i$ is the precision at top $i$ recommendations, $rel(i)$ is a binary indicator which signals if the $i^{th}$ recommended item is relevant or not and $|U|$ is the total number of users in the test set.\\ 
    
    \item \ac{NDCG} is a utility-based ranking measure which considers the order of recommended items in the list \citep{lee2017large}. It discounts the positions of the items recommended to a user \citep{deldjoo2019movie}. This metric is chosen because in a video streaming service, users may be willing to scan all the relevant videos in the recommendation list from the beginning to the end. When the relevant videos appear at a lower ranked position the utility of recommendations is slowly penalised, since videos that are more useful for the user are highly relevant \citep{shani2011evaluating}. This metric also shows high robustness to the changes of the MovieLens-10M dataset after pre-processing \citep{tousch2019robust}. Assuming the predicted rating values for the recommendations are sorted in descending order in the recommendation list for user $u$, $DCG_u$ is defined as \citep{deldjoo2019movie, matveeva2006high}
    
    \begin{equation}
        DCG_u@N = \sum_{i=1}^{N} \frac{2^{r_{u,i}} - 1}{log_2(i+1)},
    \end{equation}

    where $r_{u,i}$ is the true rating of user $u$ to the item ranked at position $i$. \ac{NDCG} is the normalised $DCG_u$ which is the ratio of $DCG_u$ to the ideal discounted cumulative gain ($IDCG_u$), which is the value that represents the ideal ranking for user $u$ calculated using the ground-truth ranking instead of the predicted one. This is computed as \citep{deldjoo2019movie}
    
    \begin{equation}
        NDCG_u = \frac{DCG_u}{IDCG_u}.
    \end{equation}
    
    The overall \ac{NDCG} is obtained by calculating the mean over the whole set of users in the test set \citep{deldjoo2019movie}.\\
\end{enumerate}

\subsubsection{Beyond-accuracy metrics}
Evaluating the recommendation generated using the various deep learning features solely according to accuracy is not sufficient since the objective of a recommender system is not only restricted to generate relevant recommendation lists to the users. Instead, the features should also cover the whole set of preferences of the users, given the huge body of video data \citep{silveira2019good}. Beyond-accuracy metrics are used to help to assess the quality of the various deep learning features explored in this work by capturing the coverage and diversity of recommendations. These metrics assess if the systems using these features are able leverage the whole catalogue instead of only a few highly popular items \citep{deldjoo2019movie}. It also assesses if the recommendation lists generated by the system for different users are being diversified. In this work, the video recommendation system using the various video content features is evaluated using the following measures:\\

\begin{enumerate}

    \item Intra-list diversity is a metric which measures the efficiency of the recommender to generate recommendation lists that cover the entire set of preferences of the users \citep{silveira2019good}. It is chosen because recommendation lists with similar items may not be of interest of the user \citep{silveira2019good}. It is calculated by using the cosine similarity between the items recommended based on genre features as \citep{deldjoo2019movie, silveira2019good}
    
          \begin{equation}
        IntraL(L) = \frac{\sum_{i\epsilon L} \sum_{j\epsilon L\setminus i}(1 - cossim(i,j))}{|L|\cdot (|L| - 1)}
    \end{equation}
    
    \begin{equation}
    cossim(i,j) = \frac{\vec{f}_i \cdot \vec{f}_j}{||\vec{f}_i||\mbox{ }||\vec{f}_j||},
    \label{eq:cosine_similarity}
    \end{equation}
    
    where $|L|$ is the length of the recommendation list $L$, $cossim(i,j)$ is the cosine similarity between items $i$ and $j$, and $\vec{f}_i, \vec{f}_j \in \mathbb{R}^{|F|}$  are the feature vectors of items $i$ and $j$ with $|F|$ number of features, respectively. Recommendation lists that contains items very similar to one another in terms of their genres obtain low values for this metric.\\

    \item Item coverage of a recommender system is the ratio of distinct items for which the system is able to make recommendations \citep{silveira2019good}. This metric is chosen because it measures the proportion of items in the catalogue that have been recommended at least once over the number of potential items. If a recommender system has low coverage it will limit the recommendations for the user thus having a direct impact on business revenue of the system and the users' satisfaction. This metric is defined as \citep{deldjoo2019movie, silveira2019good}
    \begin{equation}
        coverage = \frac{|\hat{I}|}{|I|},
    \end{equation}
    
    where $|I|$ is the total number of items in the test set catalogue and $|\hat{I}|$ is the number of items in $I$ recommended at least once by the system.\\

    \item Shannon Entropy is a measure that provides an overview of the recommender system as a whole by measuring the distributional inequality of recommendations across all users \citep{deldjoo2019movie}. This metric is chosen to better understand the capability of each deep learning feature to generate unequally different video recommendations within a certain item coverage value over the whole set of users. Shannon Entropy is defined as \citep{deldjoo2019movie}
    
    \begin{equation}
        SE = -\sum_{i\epsilon I}\frac{rec(i)}{rec_t} \cdot ln\frac{rec(i)}{rec_t},
        \label{entropy}
    \end{equation}

    where $I$ is the set of items in the scenario being evaluated, $rec(i)$ is the number of times item $i$ has been recommended across all users, $rec_t$ is the total number of recommendations. As can be seen in this equation, the Shannon entropy has a value range between $0$ and $ln(n)$ that represents when one item is recommended many times and when $n$ items are recommended equally frequently \citep{shani2011evaluating}.

\end{enumerate}

\subsection{Recommendation performance in warm-start scenario}
In this section, we report the performance of the video recommendation system in the item warm-start scenario when using different features on their content description. Table \ref{tab:warmstart_map_ndcg} reports the MAP and NDCG results at different cut-off values for the CER and scaled-CER recommender models using genre, hand-crafted and deep learning features.

	\begin{table}[H]
\caption{Results for the CER and scaled-CER recommender models with regards to accuracy metrics in the item warm-start scenario. The performance of different video content features is evaluated using the two recommender models. The highest results across the same metric are marked in bold.}
  \label{tab:warmstart_map_ndcg}
  \centering
    \begin{adjustbox}{width=\columnwidth}
    \begin{tabular}{|c|c c c c c c|}
        \cline{1-7}
Recommender models		& MAP@5 	& NDCG@5 	& MAP@15 	& NDCG@15 	& MAP@30 	& NDCG@30 	\\
\midrule
\underline{\textit{CER}} & & & & & & \\
Genres 	&0.1101	&0.1447	&0.1224	&0.2102	&0.1332	&0.2515	\\ 
\textit{Obj(IN)} 	&0.1110	&0.1458	&0.1233	&0.2111	&0.1341	&0.2523	\\
Scene	&0.1107	&0.1452	&0.1229	&0.2104	&0.1336	&0.2516	\\ 
\textit{Action(IG)} 	&0.1107	&0.1453	&0.1232	&0.2108	&0.1340	&0.2522	\\ 
\textit{Action(HMDB)} 	&0.1113	&0.1459	&0.1236	&0.2114	&0.1344	&0.2526	\\ 
iDT 	&0.1105	&0.1450	&0.1227	&0.2102	&0.1335	&0.2515	\\ 
MoSIFT 	&0.1111	&0.1457	&0.1233	&0.2110	&0.1341	&0.2520	\\ 
MFCC 	&0.1112	&0.1459	&0.1232	&0.2108	&0.1339	&0.2519	\\ 
\textit{VGGish} 	&0.1105	&0.1451	&0.1227	&0.2103	&0.1335	&0.2517	\\ 
\midrule
\underline{\textit{scaled-CER}} & & & & & &\\
Genres 	&0.1531	&0.1841	&0.1562	&0.2538	&0.1665	&0.2962	\\ 
\textit{Obj(IN)} 	&\textbf{0.1536}	&\textbf{0.1846}	&\textbf{0.1568}	&\textbf{0.2546}	&\textbf{0.1671}	&\textbf{0.2971}	\\ 
Scene 	&0.1531	&0.1841	&0.1562	&0.2539	&0.1665	&0.2961	\\ 
\textit{Action(IG)} 	&0.1530	&0.1840	&0.1562	&0.2538	&0.1665	&0.2962	\\ 
\textit{Action(HMDB)} 	&0.1533	&0.1843	&0.1564	&0.2543	&0.1668	&0.2968	\\ 
iDT 	&\textbf{0.1536}	&\textbf{0.1846}	&0.1567	&\textbf{0.2546}	&\textbf{0.1671}	&\textbf{0.2971}	\\ 
MoSIFT	&0.1535	&\textbf{0.1846}	&0.1566	&0.2545	&0.1670	&0.2970	\\ 
MFCC 	&0.1529	&0.1840	&0.1561	&0.2538	&0.1664	&0.2962	\\ 
\textit{VGGish} 	&0.1531	&0.1842	&0.1563	&0.2541	&0.1666	&0.2966	\\  \hline
    \end{tabular}
    \end{adjustbox}
\end{table}

As can be seen in Table \ref{tab:warmstart_map_ndcg}, the CER model obtained the highest results in terms of MAP and NDCG at all cut-off values when using the \textit{Action(HMDB)} features and the lowest results are obtained using genre features. However, it can be observed that the CER model exhibits similar performance across all types of video content features. This outcome is similar to \citep{du2018personalized} and suggests that in the item warm-start scenario, the interactions collected for the items are very important in order to obtain outstanding results, and the additional video content features help predictive performance of items with very few interactions. The importance of items prior ratings is more clear when looking at the performance of the scaled-CER model. The results show that the scaled-CER model improves over the CER model. The scaled-CER model obtained the best overall performance compared to any CER model's variant by using the matrix scaling technique presented in section \ref{sec:scaledCER}. The top results were achieved by the scaled-CER model using the \textit{Obj(IN)} features that outperforms the CER model using \textit{Action(HMDB)} by 38\%, 26.8\% and 24.3 \% in terms of MAP@5, MAP@15 and MAP@30, accordingly. In terms of NDCG@5, NDCG@15 and NDCG@30, the scaled-CER model, using \textit{Obj(IN)} features, outperforms the CER model using \textit{Action(HMDB)} features by 26.5\%, 20.4\% and 17.6\%, respectively. These outcomes clearly show that in the item warm-start scenario, the item content descriptor is not as important as the ratings of the items. Moreover, the results clearly illustrates the effectiveness of the matrix scaling technique, where the scaled-CER recommdender model presents the best capability to generate recommendation lists that contain relevant items at the top positions. In addition, similar to the CER model, the scaled-CER model presents similar results along the different types of video content features, however, the difference between the results is extremely small after proper scaling. 

Nevertheless, by a closer inspection of the scaled-CER model variants, it can be seen that the state-of-the-art iDT feature vectors is the best baseline video content feature and outperforms almost all the deep learning features with the only exception being the \textit{Obj(IN)} features. The \textit{VGGish} features outperforms the hand-crafted MFCC features. However, as pointed out above, the difference between the results of any type of video content feature is not significant. Furthermore, it is worth pointing out that for each model as the cutoff value increases, MAP and NDCG results also increase. This is understandable because as the number of items being recommended increases, the more likely one of them to be a true label, which means that more correctly predicted videos are obtained.

\begin{table}[H]
\caption{Results for the CER and scaled-CER recommender models with regards to beyond-accuracy metrics in the item warm-start scenario. The performance of different video content features is evaluated using the two recommender models. The highest results across the same metric are marked in bold.}
  \label{tab:warmstart_diversity}
  \centering
    \begin{adjustbox}{width=\columnwidth}
    \begin{tabular}{|c|c c c c c c c c c|}
        \cline{1-10}
Recommender models	& \begin{tabular}{@{}c@{}}Div. \\ SE\end{tabular}@5 	& \begin{tabular}{@{}c@{}}Div. \\ IntraL\end{tabular}@5 	& \begin{tabular}{@{}c@{}}Item \\ Cov.\end{tabular}@5 	& \begin{tabular}{@{}c@{}}Div. \\ SE\end{tabular}@15 	& \begin{tabular}{@{}c@{}}Div. \\ IntraL\end{tabular}@15 	& \begin{tabular}{@{}c@{}}Item \\ Cov.\end{tabular}@15 	& \begin{tabular}{@{}c@{}}Div. \\ SE\end{tabular}@30 	& \begin{tabular}{@{}c@{}}Div. \\ IntraL\end{tabular}@30 	& \begin{tabular}{@{}c@{}}Item \\ Cov.\end{tabular}@30 	\\
\midrule
\underline{\textit{CER}} & & & & & & & & &\\
Genres 	&\textbf{8.0829}	&0.4804	&\textbf{0.1568}	&\textbf{8.6157}	&0.5992	&0.2092	&\textbf{9.0273}	&0.6383	&0.2561	\\ 
\textit{Obj(IN)}  	&8.0806	&0.4899	&0.1540	&8.6059	&0.6076	&0.2007	&9.0152	&0.6455	&0.2373	\\ 
Scene 	&8.0763	&0.4895	&0.1548	&8.6076	&0.6069	&0.2007	&9.0184	&0.6452	&0.2383	\\
\textit{Action(IG)} 	&8.0791	&0.4883	&0.1558	&8.6102	&0.6067	&0.2016	&9.0209	&0.6448	&0.2387	\\ 
\textit{Action(HMDB)} 	&8.0791	&0.4897	&0.1538	&8.6090	&0.6071	&0.1999	&9.0186	&0.6451	&0.2364	\\ 
iDT 	&8.0725	&0.4901	&0.1542	&8.6051	&0.6073	&0.2001	&9.0164	&0.6455	&0.2369	\\ 
MoSIFT 	&8.0755	&0.4905	&0.1543	&8.6062	&0.6070	&0.1999	&9.0195	&0.6451	&0.2359	\\ 
MFCC	&8.0725	&\textbf{0.4913}	&0.1539	&8.6080	&\textbf{0.6076}	&0.1995	&9.0198	&\textbf{0.6458}	&0.2362	\\ 
\textit{VGGish} 	&8.0673	&0.4897	&0.1538	&8.6009	&0.6063	&0.2014	&9.0088	&0.6448	&0.2396	\\ 
\midrule
\underline{\textit{scaled-CER}} & & & & & & & & &\\
Genres 	&7.5366	&0.4707	&0.1480	&8.3193	&0.5883	&\textbf{0.2283}	&8.8602	&0.6277	&\textbf{0.3096}	\\ 
\textit{Obj(IN)} 	&7.5225	&0.4777	&0.1372	&8.2966	&0.5953	&0.2055	&8.8316	&0.6338	&0.2669	\\
Scene 	&7.5295	&0.4788	&0.1398	&8.3052	&0.5954	&0.2092	&8.8406	&0.6337	&0.2742	\\ 
\textit{Action(IG)}	&7.4964	&0.4774	&0.1404	&8.2807	&0.5954	&0.2101	&8.8209	&0.6342	&0.2738	\\ 
\textit{Action(HMDB)} 	&7.4668	&0.4776	&0.1385	&8.2552	&0.5957	&0.2103	&8.7996	&0.6344	&0.2746	\\ 
iDT	&7.5018	&0.4765	&0.1361	&8.2871	&0.5954	&0.2040	&8.8252	&0.6346	&0.2632	\\ 
MoSIFT 	&7.5162	&0.4783	&0.1365	&8.2947	&0.5962	&0.2025	&8.8288	&0.6351	&0.2609	\\ 
MFCC 	&7.4987	&0.4787	&0.1375	&8.2777	&0.5969	&0.2051	&8.8171	&0.6355	&0.2665	\\ 
\textit{VGGish} 	&7.4990	&0.4777	&0.1387	&8.2796	&0.5957	&0.2110	&8.8190	&0.6341	&0.2759	\\ \hline
    \end{tabular}
    \end{adjustbox}
\end{table}

Table \ref{tab:warmstart_diversity} presents the beyond-accuracy performance of the CER and scaled-CER recommender models using genre, hand-crafted and deep learning features. The performance is measured in terms of Shannon Entropy, intra-list diversity and item coverage measured at different cut-off values. The CER model achieved the highest results in almost all the metrics with the exception of item coverage at cut-off values 15 and 30. This outcome was expected because it is known that there exists an inherent trade-off between accuracy and beyond-accuracy metrics \citep{adomavicius2011improving}. The matrix scaling technique used by the scaled-CER model brings great improvements in terms of accuracy metrics at a cost of intra-list diversity and Shannon Entropy. It is interesting to note that for the cut-off values 15 and 30, the highest results for item coverage were obtained by the scaled-CER model. This outcome is in line with the main purpose of the scaling factor which is to increase the sensitivity of the model to rare items. However, the increase in item coverage is accompanied by less diverse recommendation lists where a number of items were recommended more times than the other items. 

By a closer inspections of the video content features, the CER model using genre features obtained the highest results for Shannon Entropy for all cut-off values and the highest item coverage for cut-off value 5. Take into consideration the fact that this variant of the CER, obtained the worst performance in terms of accuracy metrics. In terms of intra-list diversity results, the CER model using genre features obtained the lowest results compared to other CER variants. This outcome was expected because genre features are used to calculate this metric. The results suggest that the model is generating recommendation lists with a number of videos of the same genre. Moreover, it is interesting to note that the scaled-CER model using \textit{Obj(IN)} features did not obtained the lowest results in terms of beyond-accuracy metrics since this model obtained the best performance in terms of accuracy metrics. In addition, similar to accuracy metric results, the beyond-accuracy results for the CER and scaled-CER models are similar for the various video content features along the respective cut-off values, and they increase with an increase in length of the recommendation list.

\subsection{Recommendation performance in cold-start scenario}\label{sec:recommendation_cold_start_scenario_experiment}
The experiments results of the video recommendation system in the new item cold-start scenario are presented in this section. The system is evaluated when using each feature explored in this work on its content description. The results obtained in this scenario represent the ability of the system to alleviate the new item cold-start problem.

	\begin{table}[H]
\caption{Results for the CER and scaled-CER recommender models with regards to accuracy metrics in the item cold-start scenario. The performance of different video content features is evaluated using the two recommender models. The highest results across the same metric are marked in bold.}
  \label{tab:coldstart_map_ndcg}
  \centering
    \begin{adjustbox}{width=\columnwidth}
    \begin{tabular}{|c|c c c c c c|}
        \cline{1-7}
Recommender models		& MAP@5 	& NDCG@5 	& MAP@15 	& NDCG@15 	& MAP@30 	& NDCG@30 	\\
\midrule
\underline{\textit{CER}} & & & & & & \\
Genres 	&0.0099	&0.0141	&0.0112	&0.0268	&0.0130	&0.0386	\\ 
\textit{Obj(IN)}	&0.0159	&0.0223	&0.0170	&0.0367	&0.0189	&0.0497	\\
Scene 	&0.0152	&0.0210	&0.0156	&0.0334	&0.0172	&0.0445	\\ 
\textit{Action(IG)} 	&0.0146	&0.0201	&0.0150	&0.0331	&0.0166	&0.0447	\\ 
\textit{Action(HMDB)} 	&0.0136	&0.0185	&0.0140	&0.0309	&0.0155	&0.0424	\\ 
iDT 	&0.0098	&0.0137	&0.0102	&0.0234	&0.0114	&0.0326	\\ 
MoSIFT 	&0.0095	&0.0131	&0.0096	&0.0215	&0.0106	&0.0297	\\ 
MFCC 	&0.0111	&0.0155	&0.0117	&0.0258	&0.0129	&0.0350	\\ 
\textit{VGGish} 	&0.0134	&0.0187	&0.0139	&0.0307	&0.0154	&0.0414	\\ 
\midrule
\underline{\textit{scaled-CER}} & & & & & &\\
Genres 	&0.0113	&0.0159	&0.0125	&0.0294	&0.0143	&0.0416	\\ 
\textit{Obj(IN)} 	&\textbf{0.0178}	&\textbf{0.0247}	&\textbf{0.0188}	&\textbf{0.0404}	&\textbf{0.0208}	&\textbf{0.0538}	\\
Scene &0.0162	&0.0221	&0.0165	&0.0350	&0.0181	&0.0463	\\ 
\textit{Action(IG)} 	&0.0159	&0.0215	&0.0160	&0.0345	&0.0176	&0.0459	\\ 
\textit{Action(HMDB)} 	&0.0137	&0.0188	&0.0140	&0.0311	&0.0155	&0.0422	\\ 
iDT 	&0.0114	&0.0157	&0.0117	&0.0264	&0.0130	&0.0362	\\ 
MoSIFT 	&0.0105	&0.0143	&0.0105	&0.0234	&0.0116	&0.0321	\\ 
MFCC 	&0.0116	&0.0163	&0.0121	&0.0266	&0.0134	&0.0360	\\ 
\textit{VGGish} 	&0.0138	&0.0193	&0.0144	&0.0319	&0.0160	&0.0430	\\ \hline
    \end{tabular}
    \end{adjustbox}
\end{table}

In contrast to the item warm-start scenario, it can be seen in Table \ref{tab:coldstart_map_ndcg} that in the item cold-start scenario, the CER and scaled-CER models exhibit results that are more varied across different types of video content features. This suggests that these models are relying more on the features to generate recommendations and each type of video content feature discriminates the user preferences differently. Similar to the item warm-start scenario, the scaled-CER model achieved the highest results with regards to all the accuracy metrics and a noticeable improvement over the CER model is observed. This outcome shows the effectiveness of the matrix scaling technique in the item cold-start scenario as well, which is able to improve the performance of different types of features. It suggests that the collaborative information learnt with the item popularity sensitivity adjustment along with video content features is very important to recommend cold items with high precision. It can be observed that the scaled-CER, using MFCC features, it is the best baseline with regards to the cut-off value 5 across the respective metrics, however with regards to the cut-off values 15 and 30, the best baseline is the genre features. The best overall performance is obtained by Obj(IN) features which outperform the MFCC features by 53.4\% and 51.5\% in terms of MAP@5 and NDCG@5, accordingly. In terms of MAP@15, MAP@30, NDCG@15 and NDCG@30, the Obj(IN) features outperforms genre features by 50.4\%, 45.4\%, 37.4\% and 29.3\%, respectively. These results indicate that the scaled-CER model using Obj(IN) features in its content descriptor provides considerable better recommendations that are placed at the top of the recommendation list compared to the genre and hand-crafted features. In addition, as the recommendation list gets longer the more relevant items the model is able to recommend.  

It is interesting to note that action-centric deep learning features present noticeably better performance compared to the hand-crafted MoSIFT features and state-of-the-art hand-crafted iDT features across all accuracy metrics. The best action features with regards to MAP and NDCG across different cut-off values is \textit{Action(IG)} followed by \textit{Action(HMDB)}. These results are very promising since it shows that the motion information captured by deep learning features lead to better recommendations compared to the hand-crafted iDT and MoSIFT features in terms of accuracy metrics. 

Similarly, deep learning audio features outperform the hand-crafted MFCC features. This confirms the success of deep learning features in the video recommendation context in terms of accuracy metrics. In addition, it can also be noted that all the deep learning features explored in this work outperforms genre features which thus emphasises the importance of using non-textual features extracted from videos to improve cold item recommendations.

		\begin{table}[H]
\caption{Results for the CER and scaled-CER recommender models with regards to beyond-accuracy metrics in the new item cold-start scenario. The performance of different video content features is evaluated using the two recommender models. The highest results across the same metric are marked in bold.}
  \label{tab:coldstart_diversity}
  \centering
    \begin{adjustbox}{width=\columnwidth}
    \begin{tabular}{|c|c c c c c c c c c|}
        \cline{1-10}
Recommender models	& \begin{tabular}{@{}c@{}}Div. \\ SE\end{tabular}@5 	& \begin{tabular}{@{}c@{}}Div. \\ IntraL\end{tabular}@5 	& \begin{tabular}{@{}c@{}}Item \\ Cov.\end{tabular}@5 	& \begin{tabular}{@{}c@{}}Div. \\ SE\end{tabular}@15 	& \begin{tabular}{@{}c@{}}Div. \\ IntraL\end{tabular}@15 	& \begin{tabular}{@{}c@{}}Item \\ Cov.\end{tabular}@15 	& \begin{tabular}{@{}c@{}}Div. \\ SE\end{tabular}@30 	& \begin{tabular}{@{}c@{}}Div. \\ IntraL\end{tabular}@30 	& \begin{tabular}{@{}c@{}}Item \\ Cov.\end{tabular}@30 	\\
\midrule
\underline{\textit{CER}} & & & & & & & & &\\
Genres 	&7.8355	&0.2376	&0.4577	&8.5849	&0.3217	&0.6382	&9.0612	&0.3742	&0.7058	\\ 
\textit{Obj(IN)} 	&8.7835	&0.4830	&0.7851	&9.3267	&0.5876	&0.9215	&9.6660	&0.6283	&0.9683	\\
Scene 	&8.8372	&0.4949	&0.7875	&9.3987	&0.5978	&0.9115	&9.7426	&0.6355	&0.9610	\\ 
\textit{Action(IG)} 	&8.9225	&0.4740	&0.8029	&9.4278	&0.5757	&0.9179	&9.7437	&0.6152	&0.9621	\\ 
\textit{Action(HMDB)}	&9.0500	&0.4825	&0.8306	&9.5716	&0.5876	&0.9322	&9.8850	&0.6271	&0.9693	\\ 
iDT 	&9.0538	&0.4976	&0.8090	&9.5686	&0.6015	&0.9303	&9.8801	&0.6389	&0.9722	\\ 
MoSIFT	&8.1329	&0.5352	&0.5704	&8.8799	&\textbf{0.6441}	&0.7655	&9.3488	&\textbf{0.6830}	&0.8714	\\ 
MFCC	&8.6740	&\textbf{0.5450}	&0.7545	&9.3151	&0.6418	&0.9010	&9.6971	&0.6709	&0.9589	\\ 
\textit{VGGish} 	&\textbf{9.1818}	&0.4962	&\textbf{0.8777}	&\textbf{9.6664}	&0.5968	&\textbf{0.9615}	&9.9498	&0.6300	&\textbf{0.9855}	\\ 
\midrule
\underline{\textit{scaled-CER}} & & & & & & & & &\\
Genres 	&8.0557	&0.2243	&0.5136	&8.7664	&0.3087	&0.6803	&9.2270	&0.3611	&0.7442	\\ 
\textit{Obj(IN)} 	&8.6569	&0.4791	&0.7332	&9.2425	&0.5819	&0.8899	&9.6058	&0.6228	&0.9541	\\
Scene 	&8.8951	&0.4868	&0.7827	&9.4651	&0.5919	&0.9105	&9.8077	&0.6315	&0.9611	\\ 
\textit{Action(IG)} 	&9.1571	&0.4711	&0.8577	&9.6278	&0.5751	&0.9519	&9.9143	&0.6155	&0.9801	\\ 
\textit{Action(HMDB)} 	&9.1299	&0.4781	&0.8440	&9.6537	&0.5850	&0.9470	&\textbf{9.9591}	&0.6254	&0.9790	\\ 
iDT 	&8.8032	&0.4874	&0.7342	&9.3868	&0.5919	&0.8888	&9.7340	&0.6304	&0.9494	\\ 
MoSIFT 	&7.9336	&0.5308	&0.4944	&8.7381	&0.6379	&0.7003	&9.2389	&0.6772	&0.8167	\\ 
MFCC 	&8.4761	&0.5417	&0.6790	&9.1700	&0.6390	&0.8496	&9.5810	&0.6676	&0.9308	\\ 
\textit{VGGish} 	&9.1073	&0.4888	&0.8656	&9.6207	&0.5897	&0.9561	&9.9195	&0.6234	&0.9834	\\ \hline
    \end{tabular}
    \end{adjustbox}
\end{table}

Table \ref{tab:coldstart_diversity} reports the beyond-accuracy results of the CER model and scaled-CER model in terms of Shannon Entropy, intra-list diversity and item coverage in the item cold-start scenario. Overall, the CER model obtained the highest results for almost all three metrics with the exception being the SE@30. However, these results come at the cost of accuracy. The lowest results are obtained using the genre features. 

In contrast to the item warm-start scenario, in the item cold-start scenario, the scaling factor of the scaled-CER model does not lead to the highest results with regards to item coverage. The scaling factor only increases the item coverage of the genre, \textit{Action(IG)} and \textit{Action(HMDB)} features. It is interesting to see that this increase comes with an increase in Shannon Entropy but with a slight decrease in intra-list diversity. This means that the number of items that are recommended equally often increased, however, the number of recommendations of the same genre also increased. In addition, it is important to note that the action-centric deep learning features present a better item coverage with recommendations that are harder to guess in comparison to the hand-crafted iDT and MoSIFT features without compromising recommendation accuracy. A similar outcome is observed for the deep learning audio features compared to MFCC features. Furthermore, it is also worth mentioning that the best visual-appearance, action and audio features in terms of MAP and NDCG, namely \textit{Obj(IN)}, \textit{Action(IG)} and \textit{VGGish} features, obtained item coverage results in the range of 0.7332 to 0.8656 for the smallest cut-off value and 0.9541 to 0.9834 for the highest cut-off value. \textit{These results are promising and suggest that the scaled-CER model is able to recommend more than 95\% of cold items and these items are highly relevant to users.} 

\subsection{Evaluation of different fusion methods}
The evaluation of different fusion methods is performed in the item cold-start scenario. The goal of this experiment is to address the problem: Given videos represented by multiple features, namely visual-appearance, audio and action features, how should we further improve the recommendation of newly added videos. The experiment is based on the combination of the most accurate deep learning feature modalities, namely visual-appearance, audio and action features, reported in section \ref{sec:recommendation_cold_start_scenario_experiment}. We use the scaled-CER model to evaluate the fusion methods described in section \ref{sec:fusionAlgorithm}. This model is chosen due to the outstanding overall performance presented in section \ref{sec:recommendation_cold_start_scenario_experiment}. The best single video content feature is used as a unimodal baseline to determine whether a fusion method really improve recommendation quality.

			\begin{table}[H]
\caption{Results of different fusion methods with respect to accuracy metrics using the best visual-appearance, audio and action features. The highest results along the respective metric are marked in bold.}
  \label{tab:fusion_MAP_NDCG}
  \centering
    \begin{adjustbox}{width=\columnwidth}
    \begin{tabular}{|c|c c c c c c c|}
        \cline{1-8}
Features	& \begin{tabular}{@{}c@{}}Feature \\ Fusion\end{tabular} & MAP@5 	& NDCG@5 	& MAP@15 	& NDCG@15 	& MAP@30 	& NDCG@30 	\\ 
\midrule
\textit{Obj(IN)} 	& - &0.0178	&0.0247	&0.0188	&0.0404	&0.0208	&0.0538	\\ 
\textit{Obj(IN) + VGGish} 	& concat &0.0215	&0.0295	&0.0223	&0.0464	&0.0244	&0.0609	\\ 
\textit{Obj(IN) + VGGish} 	& sum &0.0144	&0.0201	&0.0148	&0.0328	&0.0164	&0.0439	\\ 
\textit{Obj(IN) + VGGish} 	& max &0.0122	&0.0171	&0.0126	&0.0268	&0.0137	&0.0354	\\ 
\textit{Obj(IN) + VGGish + Action(IG)} 	& concat &\textbf{0.0234}	&\textbf{0.0318}	&\textbf{0.0241}	&\textbf{0.0496}	&\textbf{0.0263}	&\textbf{0.0646}	\\ 
\textit{Obj(IN) + VGGish + Action(IG)} & sum &0.0111	&0.0156	&0.0119	&0.0277	&0.0134	&0.0386	\\ 
\textit{Obj(IN) + VGGish + Action(IG)} 	& max &0.0086	&0.0123	&0.0091	&0.0210	&0.0102	&0.0292	\\ \hline
    \end{tabular}
    \end{adjustbox}
\end{table}

From Table \ref{tab:fusion_MAP_NDCG}, it can be seen that the sum and max fusion methods are not able to outperform the \textit{Obj(IN)} features baseline with regards to MAP and NDCG metrics across all cut-off values. An interesting observation is that the sum of \textit{Obj(IN)} and \textit{VGGish} features leads to better recommendation accuracy compared to \textit{VGGish} features alone (Table \ref{tab:coldstart_map_ndcg}). A noticeable drop in performance is observed when the \textit{Action(IG)} features are combined with \textit{Obj(IN)} and \textit{VGGish} features using the two aforementioned fusion methods. This outcome suggests that the sum and max fusion methods are not able to create a shared latent space which is easy to learn the complementary video information encoded in these features. On the other hand, it is clear that the concat fusion method outperforms the baseline, and the sum and max fusion methods with regards to all accuracy metrics across all cut-off values. The concat of \textit{Obj(IN)} and \textit{VGGish} features significantly improves upon the \textit{Obj(IN)} performance by 20.7\% and 19.4\% for the @5 cut-off experiments along the MAP and NDCG metrics, respectively. For the @15 and @30 cut-off experiments the increase over the baseline are 18.6\% and 17.3\% for MAP, and 14.8\% and 13.1\% for NDCG. In addition, different from the sum and max fusion methods, the concat of \textit{Action(IG)}, \textit{Obj(IN)} and \textit{VGGish} features presents a significant positive effect in the overall recommendation performance. More precisely, the recommendation accuracy in terms of MAP@5, MAP@15 and MAP@30 increased by 8.8\%, 8.0\% and 7.7\% over the concat of \textit{Obj(IN)} and \textit{VGGish} features, accordingly. In terms of NDCG@5, NDCG@15 and NDCG@30, the recommendation accuracy increased by 7.7\%, 6.8\% and 6.0\%, respectively. These results suggest that the shared latent space created by the concat fusion method is more discriminative thus leading to better recommendation accuracy. \textit{VGGish} features complement the recommendation accuracy. In addition, \textit{Action(IG)} features combined with \textit{Obj(IN)} and \textit{VGGish} features create video representations that are highly predictive of user preferences. This means that the content present in the videos are better described. As a result, enhanced recommendations are provided.

			\begin{table}[H]
\caption{Results of different fusion methods with respect to beyond-accuracy metrics using the best visual-appearance, audio and action features. The highest results along the respective metric are marked in bold. }
  \label{tab:fusion_diversity}
  \centering
    \begin{adjustbox}{width=\columnwidth}
    \begin{tabular}{|c|c c c c c c c c c c|}
        \cline{1-11}
Features	& \begin{tabular}{@{}c@{}}Feature \\ Fusion\end{tabular}	& \begin{tabular}{@{}c@{}}Div. \\ SE\end{tabular}@5 	& \begin{tabular}{@{}c@{}}Div. \\ IntraL\end{tabular}@5 	& \begin{tabular}{@{}c@{}}Item \\ Cov.\end{tabular}@5 	& \begin{tabular}{@{}c@{}}Div. \\ SE\end{tabular}@15 	& \begin{tabular}{@{}c@{}}Div. \\ IntraL\end{tabular}@15 	& \begin{tabular}{@{}c@{}}Item \\ Cov.\end{tabular}@15 	& \begin{tabular}{@{}c@{}}Div. \\ SE\end{tabular}@30 	& \begin{tabular}{@{}c@{}}Div. \\ IntraL\end{tabular}@30 	& \begin{tabular}{@{}c@{}}Item \\ Cov.\end{tabular}@30 	\\
\midrule
\textit{Obj(IN)} 	& - &8.6569	&0.4791	&0.7332	&9.2425	&0.5819	&0.8899	&9.6058	&0.6228	&0.9541	\\ 
\textit{Obj(IN) + VGGish} &	concat &8.9131	&0.4594	&0.8540	&9.4868	&0.5654	&0.9552	&9.8176	&0.6068	&0.9866	\\ 
\textit{Obj(IN) + VGGish} 	& sum &8.7458	&0.5012	&0.7661	&9.3273	&0.6004	&0.9056	&9.6754	&0.6349	&0.9583	\\ 
\textit{Obj(IN) + VGGish} 	& max &8.6516	&\textbf{0.5227}	&0.7293	&9.3001	&\textbf{0.6197}	&0.8858	&9.6776	&0.6497	&0.9513	\\ 
\textit{Obj(IN) + VGGish + Action(IG)} 	& concat &\textbf{8.9552}	&0.4534	&\textbf{0.8655}	&\textbf{9.5163}	&0.5578	&\textbf{0.9597}	&\textbf{9.8403}	&0.6006	&\textbf{0.9880}	\\ 
\textit{Obj(IN) + VGGish + Action(IG)} & sum 	&8.9218	&0.4869	&0.8101	&9.4826	&0.5900	&0.9355	&9.8093	&0.6277	&0.9773	\\ 
\textit{Obj(IN) + VGGish + Action(IG)} 	& max &8.6260	&0.5137	&0.6967	&9.2581	&0.6177	&0.8674	&9.6366	&\textbf{0.6547}	&0.9414	\\ \hline
    \end{tabular}
    \end{adjustbox}
\end{table}

In terms of beyond-accuracy metrics, it can be observed in Table \ref{tab:fusion_diversity} that the outstanding performance achieved by the concat of \textit{Obj(IN)}, \textit{VGGish} and \textit{Action(IG)} features with regards to accuracy metrics comes with a decrease in intra-list diversity. The concat fusion of these three features did not obtain intra-list diversity results higher than the baseline but the difference is between 3.6\% and 5.6\%. 

Surprisingly, we can see a noticeable performance improvement for the concat of \textit{Obj(IN)}, \textit{VGGish} and \textit{Action(IG)} features with regards to Shannon Entropy and item coverage. These improvements do not come at the expense of recommendation accuracy. The results are promising and indicate that the complementariness of \textit{Obj(IN)}, \textit{VGGish} and \textit{Action(IG)} features considers more items on the catalogue. These items are being given a better chance of being recommended leading to recommendations that are more equally spread out throughout all cold items.

\subsection{Ablation study}
In this experiment, we investigate the importance of different features in the overall recommendation quality. The main goal is to empirically assess the importance of using a diverse range of video content features while taking full advantage of the available features in the item cold-start scenario. Thus, the experiment is performed by combining all the video content features explored in this work.

The ablation study is based only on the scaled-CER model using the concatenation fusion method, due to the outstanding overall performance shown in the previous experiment. We remove each type of feature from the concatenation of all the features, denoted by $All/x$ where \textit{x} $\in$ \textit{\{Obj(IN), VGGish, Action(IG), Genres, Scene, Action(HMDB), MFCC, iDT, MoSIFT\}} is removed from \textit{All}. The recommendation quality is measured in terms of MAP and item coverage to understand to which extent the video recommendation system is able to explore the catalogue with high precision. 

			\begin{table}[H]
\caption{Ablation study of the importance of all video content features explored in this work in the overall recommendation quality in terms of MAP and item coverage. \textit{All/x} denotes removing \textit{x} from the concatenation of all the video content features. The best results across the respective metric are highlighted in bold. The box around the $5^{th}$ row highlights the highest MAP results obtained in this work.}
  \label{tab:ablation_study}
  \centering
    \begin{adjustbox}{width=\columnwidth}
    \begin{tabular}{|c|c c c c c c|}
        \cline{1-7}
Features & MAP@5 	& \begin{tabular}{@{}c@{}}Item \\ Coverage\end{tabular}@5 	& MAP@15 	& \begin{tabular}{@{}c@{}}Item \\ Coverage\end{tabular}@15 	& MAP@30 	& \begin{tabular}{@{}c@{}}Item \\ Coverage\end{tabular}@30 	\\ 
\midrule

\textit{All} 	&0.0344	&0.7716	&0.0348	&0.9174	&0.0377	&0.9726	\\ 
\textit{All/MoSIFT} 	&0.0337	&0.7720	&0.0341	&0.9128	&0.0370	&0.9681	\\ 
\textit{All/iDT} 	&0.0337	&0.7567	&0.0342	&0.9047	&0.0372	&0.9632	\\ 
\boxit{5.8in} \textit{All/MFCC}	&\textbf{0.0347}	&0.7468	&\textbf{0.0350}	&0.8970	&\textbf{0.0379}	&0.9595 \\ 
\textit{All/Action(HMDB)}	&0.0336	&0.7511	&0.0338	&0.9028	&0.0368	&0.9629	\\ 
\textit{All/Scene}  	&0.0332	&0.7686	&0.0336	&0.9122	&0.0366	&0.9681	\\ 
\textit{All/Genres} 	&0.0268	&\textbf{0.8281}	&0.0276	&\textbf{0.9413}	&0.0302	&\textbf{0.9778}	\\ 
\textit{All/Action(IG)} 	&0.0338	&0.7515	&0.0342	&0.9026	&0.0372	&0.9604	\\ 
\textit{All/VGGish}	&0.0325	&0.7747	&0.0328	&0.9152	&0.0356	&0.9678	\\ 
\textit{All/Obj(IN)} 	&0.0335	&0.7681	&0.0339	&0.9139	&0.0368	&0.9682	\\ \hline
    \end{tabular}
    \end{adjustbox}
\end{table}

Looking at Table \ref{tab:ablation_study}, when combining the various video content features, the overall results are really interesting. The combination brings a significant increase in recommendation accuracy that comes with a drop in item coverage. However, the item coverage achieved still better than the unimodal \textit{Obj(IN))} features (Table \ref{tab:fusion_diversity}). It can be observed that the biggest boost to recommendation accuracy and the noticeable decrease in item coverage is provided by the genre features in contrast to any other feature. When the genre features are removed from the combination of all the video content features (\textit{All/Genres}), we obtain the lowest MAP results, but the highest item coverage. The genre features complement the recommendation accuracy of the deep learning features and the hand-crafted features. This outcome implies that genre features probably remove ambiguity from the non-textual features that in turn lead to an improvement in MAP while sacrificing item coverage. This was expected because each non-textual content feature vector was fused with genres that describe high-level concepts of a movie thus creating a more discriminative semantically meaningful content descriptor. By removing any other feature, we can see a decrease in recommendation accuracy, for each type of feature with the only exception being the MFCC features. When MFCC features are removed from the combination of all the video content features (\textit{All/MFCC}) MAP results slightly above \textit{All} are obtained for all cut-off values. \textit{These MAP results are the highest achieved in this experiment}. However, removing MFCC features also provides a noticeable drop in item coverage for all cut-off values which are the lowest item coverage obtained in this experiment. This means that MFCC features provide a good balance between MAP and item coverage when combined with the other features explored in this work.    

Nevertheless, the combination of all the various video content features provides recommendations that are very precise but about 22.84\% of the total number of cold items are never recommended to a user for the lowest cut-off value. However, for cut-off values 15 and 30, more than 90\% of cold items are recommended to users meaning that the descriptors obtained from the combination of all the features are highly predictive of the user preferences leading to a wide range of relevant video recommendations. This shows the strong correlation between the deep learning features, hand-crafted features and especially genre features in the overall recommendation quality.

\section{Conclusion and future work}\label{sec:conclusion}

In this paper, we investigate multiple video content features, to solve the new item cold-start problem. Various deep learning features extracted from the multi-modal, extremely high dimensional information from the videos are used to enhance the quality of recommendations. The features capture visual-appearance, audio and motion information from the media contained in the videos. A comparison between these features is performed using a hybrid recommender model, namely the CER recommender model. In addition, we propose an improvement for this model using a known matrix scaling technique. The proposed improved model is named the scaled-CER. This model is sensitive to rare items by scaling the collaborative information before training.  

It is found that in the item warm-start scenario the video content information captured by the different features does not seem to be important in the recommendation performance achieved. A noticeable boost to recommendation accuracy is achieved by the matrix scaling technique. The scaled-CER obtained the best recommendation accuracy and an improvement in item coverage. However, there is a decrease in Shannon Entropy and intra-list diversity.

In the new item cold-start scenario, the various video content features are as important as the matrix scaling technique to achieve outstanding recommendation quality. Overall, the video content information captured by the deep learning features are more discriminative compared to hand-crafted features and genre features. In particular, the motion information captured by action-centric deep learning features extracted with 3D-CNNs are better than hand-crafted action features. They lead to better recommendation accuracy and item coverage where the recommendations are more balanced. For this reason, it can be concluded that the success of 3D-CNN features on tasks like action recognition and video classification also occur in the video recommendation context. A similar outcome is observed for deep learning audio features in comparison to MFCC features.

Moreover, we investigate different fusion methods to effectively combine the features before training the model, in order to improve the recommendation quality in terms of accuracy and beyond-accuracy metrics. The best fusion method was found to be the concatenation method. The results suggest that fusion of visual, audio and action features provide more accurate video recommendations to users when compared to the fusion of only visual and audio features. Apart from intra-list diversity measure, an improvement upon the fusion of visual and audio features is also observed for all the beyond-accuracy measures. 

Furthermore, an ablation study is performed where all the video content features explored in this study are combined. The results of the ablation study demonstrated that apart from one hand-crafted feature (MFCC features), all types of features, namely genre, hand-crafted and deep learning features are necessary to achieve the highest performance observed in this work in terms of accuracy. However, MFCC features provide a worthy balance between accuracy and item coverage. The results also showed that genre features are the most important features in the overall result since the largest drop is observed when they are removed from the combination with all other features. However, the high precision comes with a decrease in item coverage. 

Finally, it is worth mentioning that the proposed work has a few limitations. In the cold-start scenario, the quality of the pre-computed video features have an impact on the recommendations. The current model does not leverage user metadata which is important when dealing with cold-start users. Lastly, in order to make recommendations based on new user-video interaction data, a full model retraining is necessary to refresh the model due to the static nature of the user and video embedding matrices.

As future work, it will be worth evaluating the features used in this research study in terms of user's quality perception. \textit{This can be achieved by deploying the recommender model in a web application that serves the pre-computed recommendations to users. Then, the users will be prompt to answer a list of questions that measure the perceived quality of the recommendations}. The outcome of this experiment could assist in better tuning of the video recommendation system in industrial applications and in the creation of high quality recommendation explanations to increase trust in the system. In addition, it would be worth investigating the correlation between multimedia features, electronic programming guide (EPG) information and the user feedback gathered from the use of the remote control. The end goal would be to enhance existing recommender systems in this domain to provide more significant TV program recommendations to users that lead to a decrease in the use of the remote control while improving their experience.

\section*{Acknowledgements}

This work was funded by the MultiChoice Research Chair of Machine Learning at the University of Pretoria, South Africa.

\bibliography{mybibfile}

\end{document}

%% file: acro_list.tex
\renewcommand*{\acsfont}[1]{\normalfont#1}
\begin{acronym}[labelindent=1000cm]
\acro{NNs}{neural networks}
\acro{DNNs}{deep neural networks}
\acro{DNN}{deep neural network}
\acro{NN}{neural network}
\acro{RNN}{recurrent neural network}
\acro{LSTM}{long short-term memory}
\acro{CNN}{convolutional neural network}
\acro{SVMs}{support vector machines}
\acro{SVM}{support vector machine}
\acro{MFCCs}{Mel-frequency cepstral coefficients}
\acro{MFCC}{Mel-frequency cepstral coefficients}
\acro{LSTMs}{Long Short-Term Memory Units}
\acro{CNNs}{convolutional neural networks}
\acro{CF}{collaborative filtering}
\acro{CB}{content based}
\acro{k-NN}{k-nearest neighbor}
\acro{FM}{factorisation machine}
\acro{NDCG}{Normalised discounted cumulative gain}
\acro{MAP}{Mean average precision}
\acro{VLAD}{vectors of locally aggregated descriptors}
\acro{FV}{Fisher vectors}
\acro{PCA}{principal component analysis}
\acro{t-SNE}{t-distributed stochastic neighbour embedding}
\acro{OCR}{optical character recognition}
\acro{RMS}{root mean square}
\acro{ZCR}{zero crossing rate}
\acro{DCT}{discrete cosine transform}
\acro{SIFT}{scale-invariant feature transform}
\acro{SURF}{speeded-up robust feature}
\acro{HOG}{histograms of oriented gradient}
\acro{MPEG}{moving picture experts group}
\acro{HOF}{histogram of optical flows}
\acro{iDT}{improved dense trajectories}
\acro{STIPs}{spatio-temporal interest points}
\acro{MoSIFT}{motion scale-invariant feature transform}
\acro{I3D}{inflated 3D ConvNets}
\acro{HMDB}{human motion database}
\acro{AVSlowFast}{audiovisual SlowFast}
\acro{SSL}{self-supervised learning}
\acro{GMM}{Gaussian mixture model}
\acro{GRU}{gated recurrent unit}
\acro{CCA}{canonical correlation analysis}
\acro{URM}{user rating matrix}
\acro{ICM}{Item content matrix}
\acro{UCM}{User content matrix}
\acro{MAE}{mean absolute error}
\acro{MSE}{mean square error}
\acro{RMSE}{root-mean-square error}
\acro{MRR}{mean reciprocal ranking}
\acro{ASR}{automated speech recognition}
\acro{LMTD}{Labelled Movie Trailer Dataset}
\acro{MAE}{mean absolute error}
\acro{FusedLSTM}{Fused Long short-term memory}
\acro{CBVRP}{content based video relevance prediction}
\acro{LDA}{linear discriminant analysis}
\acro{ResNet}{residual neural networks}
\acro{SCD}{scalable colour descriptor}
\acro{CSD}{colour structure descriptor}
\acro{CLD}{colour layout descriptor}
\acro{EHD}{edge histogram descriptor}
\acro{HTD}{homogeneous texture descriptor}
\acro{cSLIM}{collective sparse linear method}
\acro{CFeCBF}{collaborative-filtering-enriched content-based filtering}
\acro{AVFs}{Aesthetic-visual features}
\acro{CER}{collaborative embedding regression}
\acro{WMF}{weighted matrix factorisation}
\acro{MBH}{motion boundary histograms}
\acro{STFT}{short-time Fourier transform}
\acro{FFT}{fast Fourier transform}
\acro{DFT}{discrete Fourier transform}
\acro{MAD}{Median absolute deviation}
\acro{IQR}{Interquartile range}
\acro{SSR}{signed square root}
\acro{ItemKNN-CBF}{Item-based k-nearest neighbors content-based filtering}
\acro{TF-IDF}{term frequency-inverse document frequency}
\acro{BM25}{Okapi best matching 25}
\acro{SVD}{singular value decomposition}
\acro{UMAP}{uniform manifold approximation and projection}
\acro{CV}{cross-validation}
\acro{VGG}{Visual Geometry Group}
\acro{RBF}{radial basis function}
\acro{REC}{recall}
\acro{Div.}{diversity}
\acro{IntraL}{intra-list}
\acro{InterL}{inter-list}
\acro{Cov.}{coverage}
\acro{SE}{Shannon entropy}
\acro{HHI}{Herfindahl-index}
\acro{concat}{concatenation}
\acro{sum}{summation}
\acro{max}{maximum}

\end{acronym}